\documentclass[final,5p,times,twocolumn,authoryear]{elsarticle}

\usepackage{graphicx}
\usepackage{dcolumn}
\usepackage{bm}
\usepackage{epsf}
\usepackage{epsfig}
\usepackage{verbatim}
\usepackage{amsmath}
\usepackage{amsfonts}
\usepackage{ifthen}
\usepackage[normalem]{ulem}
\usepackage[flushleft]{threeparttable}
\usepackage{aas_macros}
\usepackage[percent]{overpic}

\usepackage{xcolor}
\definecolor{darkgreen}{rgb}{0.0,0.5,0.0}
\usepackage[colorlinks,citecolor=darkgreen]{hyperref} 

\newcommand{\gama}{{$\gamma$}}
\newcommand{\erg}{{\mbox{ erg}}}

\newcommand{\keV}{{\mbox{ keV}}}
\newcommand{\GeV}{{\mbox{ GeV}}}
\newcommand{\kpc}{{\mbox{ kpc}}}
\newcommand{\eg}{{\emph{e.g.},}}
\newcommand{\cm}{{\mbox{ cm}}}

\newcommand{\Myr}{{\mbox{ Myr}}}

\newcommand{\GHz}{{\mbox{ GHz}}}
\newcommand{\MHz}{{\mbox{ MHz}}}

\newcommand{\coma}{{\, ,}}
\newcommand{\fin}{{\, .}}

\defcitealias{KeshetEtAl24}{K24}
\newcommand{\UK}{{\citetalias{KeshetEtAl24}}}
\defcitealias{Keshetgurwich17}{KG17}
\newcommand{\KG}{{\citetalias{Keshetgurwich17}}}

\journal{High Energy Astrophysics}

\begin{document}

\begin{frontmatter}

\title{Detection of polarized Fermi-bubble synchrotron and dust emission}

\author{Uri Keshet}
\affiliation{organization={Physics Department, Ben-Gurion University of the Negev},
            addressline={POB 653},
            city={Be'er Sheva},
            postcode={84105},
            country={Israel},
            email={; keshet.uri@gmail.com}}

\date{\today}

\begin{abstract}
The elusive polarized microwave signal from the Fermi bubbles is disentangled from the more extended polarized lobes, which similarly emanate from the Galactic plane but stretch farther west of the bubbles.
The projected $\sim20\%$ synchrotron polarization reveals magnetic fields preferentially parallel to the bubble edges, as expected downstream of a strong shock.
The projected $\sim20\%$ polarization of thermal dust emission is similarly oriented, constraining grain alignment in an extreme environment.
We argue that the larger lobes arise from an older Galactic-center, likely supermassive black-hole, outburst.
\end{abstract}

\begin{keyword}
(ISM:) cosmic rays \sep ISM: jets and outflows \sep Galaxy: center \sep Black hole physics
\end{keyword}


\end{frontmatter}

\section{Introduction}

The bipolar Fermi bubbles (FBs), emanating from the center of the Milky Way \citep{Baganoffetal03, Blandhawthorncohen03} and extending out to $|b|\gtrsim 50^\circ$ latitudes, each presents as a fairly uniform \gama-ray teardrop-shaped structure \citep{DoblerEtAl10, SuEtAl10}, demarked by an X-ray shell \citep{Keshetgurwich18}, and coincident with a dust component \citep[][henceforth \UK]{KeshetEtAl24} and with low radio frequency (\UK) to microwave \citep{Dobler12, PlanckHaze13} synchrotron emission which brightens to a haze \citep{Finkbeiner04} near the Galactic center (GC).
Their morphology, X-ray shells, and integrated \citep{SuEtAl10, Dobler12, HuangEtAl13, HooperSlatyer13, FermiBubbles14} and edge \citep[][henceforth \KG]{Keshetgurwich17} energy spectra, featuring \gama-ray and microwave cooling breaks (\UK), indicate that the FBs arose a few Myr ago, as outflows from the super-massive black hole (SMBH), which must have been collimated \citep{MondalEtAl22} nearly perpendicular to the Galactic plane \citep{SarkarEtAl23}, and identify the FB edges as strong, Mach $\gtrsim 5$ forward shocks (\KG, \UK).

The anticipated polarized signal from the FBs has drawn much attention, not only as a probe of their magnetic fields and associated physical processes, but also due to their alleged relation to pronounced, extended polarized lobes \citep{JonesEtAl12, CarrettiEtAl13} that similarly emanate from the Galactic plane, as described below.
The polarized signal from the bulk of the FBs was argued to be too weak to pick up, directly or through template decomposition, in \emph{WMAP} \citep{GoldEtAl11, Dobler12LastLook} and \emph{Planck} \citep{PlanckHaze13, PlanckDust20} maps.
However, a local linearly-polarized signal is anticipated just inside the FB edges, as the shock compresses or otherwise amplifies magnetic fields, relativistic particles, and dust.
As magnetic fields $\bm{B}$ are preferentially amplified parallel to shock fronts, a downstream polarized component with a projected position angle $\bm{\theta}\perp \bm{B}$ is expected perpendicular to the FB edges, in both synchrotron and thermal-dust emission.

Here we report a change in the polarization signature localized at the FB edges, with similar properties along different sectors along the bubbles. The signal attributed to the FBs can thus be disentangled from the smoothly varying, strong foreground and background (henceforth foreground) signals.

\section{Data and methods}

We focus on the vicinity of the FB edges as previously identified by gradient filters (\KG), and bin the polarized microwave data parallel to the edge in the same method previously used to measure the \gama-ray (\KG), X-ray \citep{Keshetgurwich18}, and radio (\UK) spectra in the near downstream.
In particular, we adopt the coarse-grained (\KG) edges of the FBs, found to effectively confine their signature in the $3$--$30\GeV$, 15-year \emph{Fermi}-LAT data used in {\UK} and reexamined here.
Non-polarized radio, microwave, infrared, and \gama-ray data are also analyzed, in the very same method, for comparison and reference.

Following {\UK}, our microwave analysis is based on the same full mission, bandpass-leakage corrected maps at $28.5,\ 44.1,\ 70.3,\ 100,\ 143,\ 217,\ 353,\ 545$, and $857$ GHz frequencies from \emph{Planck}\footnote{See \url{https://pla.esac.esa.int/\#home}}, and the nine-year, deconvolved maps of the $22.8,\ 33.2,\ 41.0,\ 61.4,$ and $94.0$ GHz bands from \emph{WMAP}\footnote{See \href{https://lambda.gsfc.nasa.gov/product/map/dr5/maps_deconv_band_r9_i_9yr_get.cfm}{https://lambda.gsfc.nasa.gov/product/map/dr5}\\
\href{https://lambda.gsfc.nasa.gov/product/map/dr5/maps_deconv_band_r9_i_9yr_get.cfm}{/maps\_deconv\_band\_r9\_i\_9yr\_get.cfm}
}.
Here, in addition to the total brightness $I$ maps analyzed in {\UK}, we also study the linearly polarized $Q$ and $U$ maps, where $I$, $Q$, and $U$ are Stokes parameters.

In order to obtain cleaner $I$ maps, needed for estimates of the polarization fraction, we first subtract their cosmic-microwave background (CMB) component, estimated using the
Planck HFI internal linear combination \cite[PILC;][]{PlanckHaze13}.
This map is constructed by a superposition of 143--545 GHz maps, designed to spectrally isolate the CMB from the dust template of \citet{FDS99}.
We mask the point sources from \emph{WMAP}'s nine-year point-source catalog and \emph{Planck}'s second catalogue of point sources (PCCS2) in the relevant bands, and the following large-scale structures: the Small and Large Magellanic Clouds, M31, NGC5090, NGC5128, Orion-Barnard Loop, and $\zeta$ Oph.
We also mask pixels where dust extinction at H$\alpha$ frequencies is larger than 1 mag,
and pixels where the H$\alpha$ intensity is greater than 10 Rayleigh.
All maps are masked at the highest, $N_{\mbox{\tiny{side}}}=2048$ HEALPix \citep{GorskiEtAl05} resolution available.

For the \gama-ray data, we use the archival Pass-8 LAT data from the Fermi Science Support Center (FSSC)\footnote{See \url{http://fermi.gsfc.nasa.gov/ssc}}, and the Fermi Science Tools (version \texttt{v10r0p5}).
Weekly all-sky files are used, spanning weeks $9$ through $789$ for a total of $781$ weeks ($\sim15$ years), with ULTRACLEANVETO class photon events.
A $90^\circ$ zenith angle cut is applied to avoid CR-generated $\gamma$-rays from the Earth's atmospheric limb, according to the appropriate
FSSC Data Preparation recommendations.
Good time intervals are selected using the recommended expression \texttt{(DATA\_QUAL==1) and (LAT\_CONFIG==1)}.
Point-source contamination is minimized by masking pixels within the $95\%$ total-event containment area of each point source in the LAT fourth source catalog \cite[4FGL;][]{AbdollahiEtAl20}.

Also used are data extracted from $45\MHz$ \citep{AlvarezEtAl97, MaedaEtAl99}, reprocessed $408\MHz$ \citep{Haslam,Remazeilles2015}, $1.4\GHz$ \citep{Reich82, ReichReich86, ReichEtAl01}, $160\,\mu\mbox{m}$ and $140\,\mu\mbox{m}$ AKARI \citep{AKARI15}, and $100\,\mu\mbox{m}$ and $60\,\mu\mbox{m}$ IRIS \citep{IRIS05} maps.

Data are binned parallel to the FB edge in the method of {\KG}, which removes smooth foregrounds in the vicinity of the edge.
Namely, we apply a non-Euclidean, signed distance transform of the data on the plane of the sky, to obtain for each pixel the oriented angular distance $\psi$ from the edge, directed outside the bubble. The result $s(\psi)$ is binned at fixed $\psi$ intervals, and a foreground model $s_f(\psi)$ is subtracted.
As we are only concerned with short, $|\psi|\lesssim 15^\circ$ distances from the edge, a constant foreground provides a reasonable approximation, so we nominally adopt a fixed $s_f(\psi)=\langle s(0<\psi<10^\circ)\rangle$, although a linear $s_f(\psi)$ fit to the external signal $s(\psi>0)$ is also considered.
A sharp transition at $\psi\simeq 0$, with the signal suddenly strengthening inward ($\psi<0$) of the predetermined edge, would then imply emission associated with the FBs, separating it from the foreground. The significance of this FB-association is evaluated using the $\mbox{TS} = \chi^2_- - \chi^2_+$ TS-test, which compares the $\mathsf{n}$ degree of freedom $\chi^2$ fit values obtained before ($-$ subscripts) and after ($+$ subscript) adding a simple, linear $s(\psi<0)$ excess to a linear model of the foreground; TS then approximately follows a chi-squared distribution $\chi_\mathsf{n}^2$ of order $\mathsf{n}\equiv \mathsf{n}_+-\mathsf{n}_-$ \citep{Wilks1938}.

\section{Results}

\begin{figure}[b!]
    \centerline{
    \includegraphics[width=1\linewidth,trim={0cm 0cm 0cm 0cm},clip]{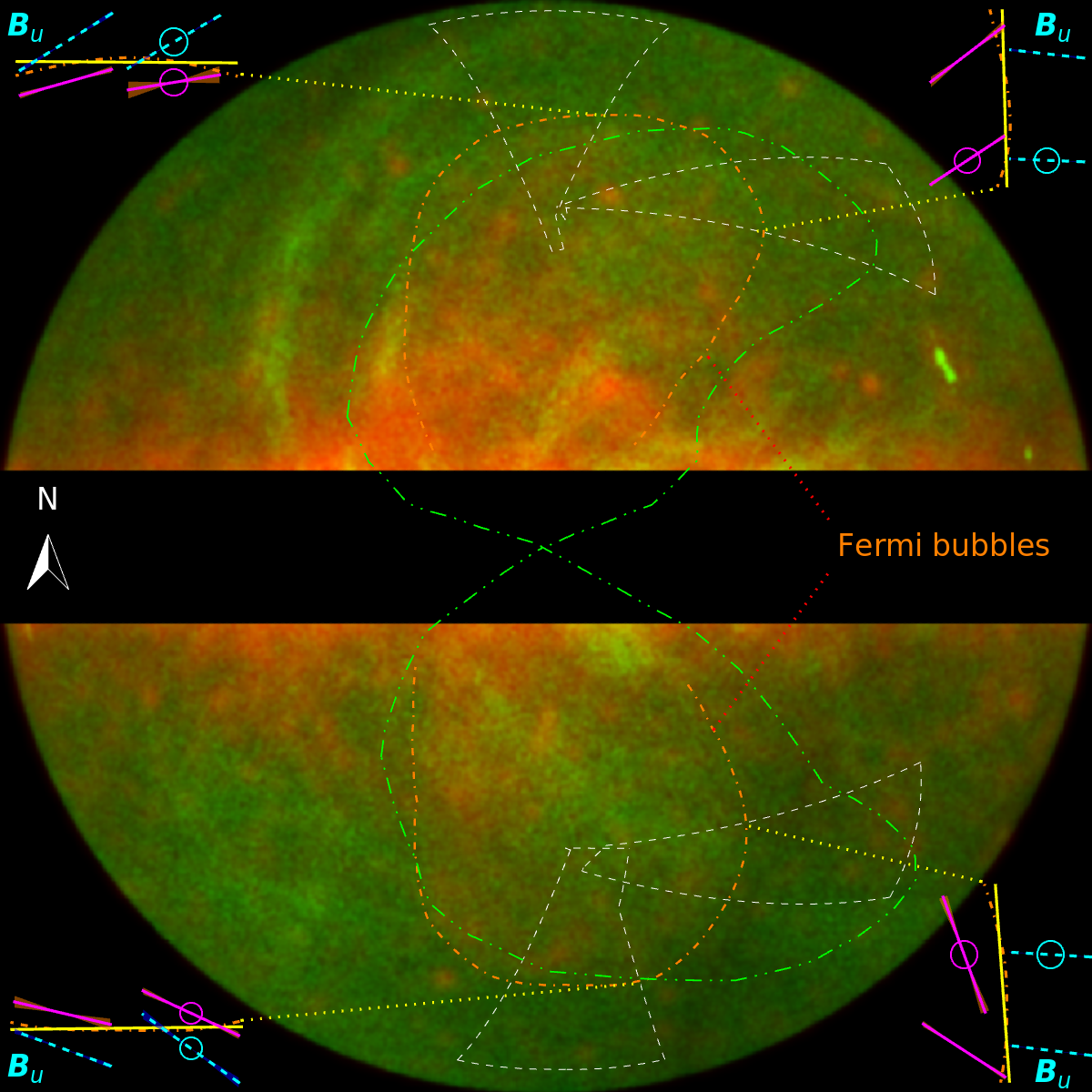}
    }
	\caption{\label{fig:RGB}
    \!\!\!False-color image of \gama-ray ($3$--$30\GeV$ \emph{Fermi}-LAT; red) and polarized microwave ($23\GHz$ \emph{WMAP}; green) intensities in an orthographic projection; the bright, $|b|\lesssim5^\circ$ Galactic plane is masked.
    The polarized lobes (with putative edges \citep{CarrettiEtAl13} shown as double-dot-dashed green curves) extend west of the FBs (edges of {\KG} in dot-dashed orange).
    For each of the four sectors analyzed (dashed white boundaries), we show (in the respective image corner) the FB edge in a Cartesian projection, along with its mean orientation (solid yellow line) and the directions of the projected upstream (cyan dashed lines with shaded dispersion) and downstream excess (solid magenta with orange dispersion) magnetic fields, as inferred from synchrotron (\emph{Planck} $30\GHz$) and dust (\emph{Planck} $353\GHz$; denoted by circles) polarizations.
	}
\end{figure}

By focusing on the predetermined FB edges, we avoid confusion with the aforementioned extended polarized lobes, observed in $23\GHz$ \emph{WMAP} \citep{JonesEtAl12} and $2.3\GHz$ S-PASS \citep{CarrettiEtAl13} data.
Figure \ref{fig:RGB} shows the \gama-ray (red) FB edges (dot-dashed orange curves, {\KG}), and the polarized synchrotron (green) putative\footnote{The lobes were not shown to sharply cut off at the purported edges, and may be more extended, than claimed by \citet{CarrettiEtAl13}.} lobe-edges \citep[double-dot dashed green curves,][]{CarrettiEtAl13}.
These polarized lobes, partly overlapping the east edges of the FBs, were tentatively identified with the bubbles, despite their different morphology.

There are, however, several reasons to doubt this lobe--FB association, and test it using the FB edges.
First, the lobes extend far beyond the west edges of the FBs, so they could not arise from the same outburst that drove the FB forward shocks.
A similar argument holds for the shear size of the lobes, spanning a solid angle $\gtrsim 40\%$ larger than the FBs.
Second, the lobes are far less east--west symmetric than the FBs, presenting a much stronger westward curvature with increasing $|b|$.
Third, the lobe ridges are traced down to the $|b|\simeq 10^\circ$ depolarization limit, where they already reach longitudes $l\gtrsim 10^{\circ}$ east of the GC \citep{CarrettiEtAl13}; therefore, the lobes either emanate far east of the GC, or emanate from the GC but initially head east before sharply curving back west at $|b|<10^{\circ}$, unlike the FBs.
Finally, the polarization orientation is approximately perpendicular to the ridges, with position angles $0\lesssim\mp\theta(\pm b\gtrsim10^{\circ})\lesssim30^{\circ}$, showing little correlation with the FBs.
Here and henceforth, $\theta$ is measured clockwise, south due east, in \emph{WMAP} rather than IAU conventions \citep{Alighieri17}, to match the \emph{WMAP} and \emph{Planck} data used.

\begin{figure*}[h!]
    \begin{center}
        \hspace{-0.3cm}
        \begin{overpic}[width=0.47\linewidth]{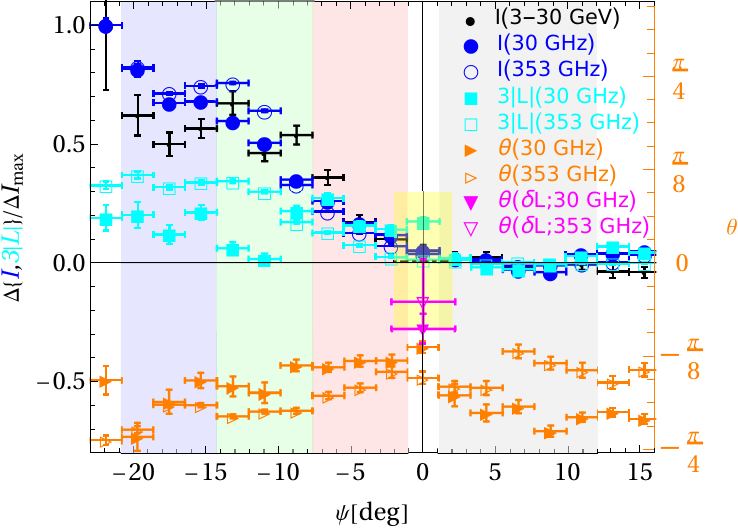}
            \put (40,73) {\small\textcolor{black}{North sector}}
        \end{overpic}
        \hspace{0.7cm}
        \begin{overpic}[width=0.47\linewidth]{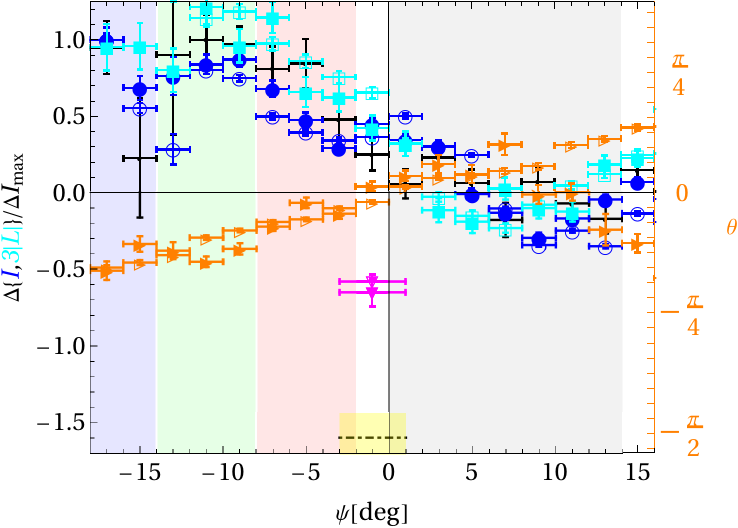}
            \put (37,73) {\small\textcolor{black}{Northwest sector}}
        \end{overpic}
    \end{center}
    \vspace{0.3cm}
    \begin{center}
        \hspace{-0.3cm}
        \begin{overpic}[width=0.47\linewidth]{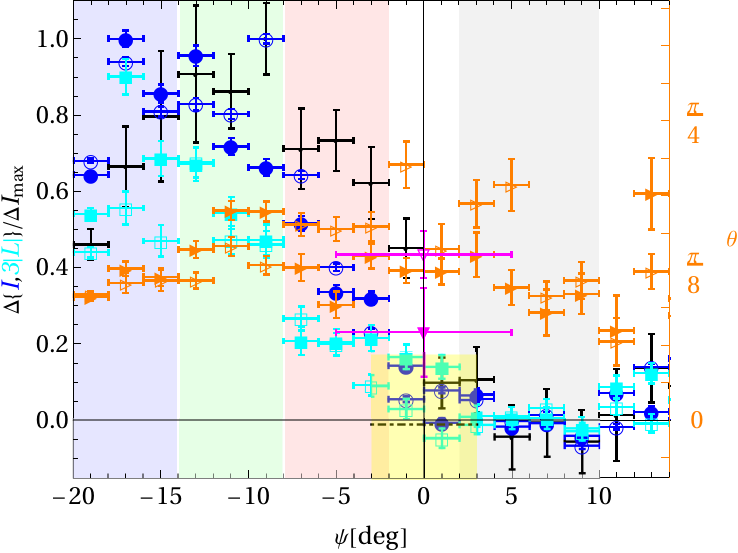}
            \put (40,77) {\small\textcolor{black}{South sector}}
        \end{overpic}
        \hspace{0.7cm}
        \begin{overpic}[width=0.47\linewidth]{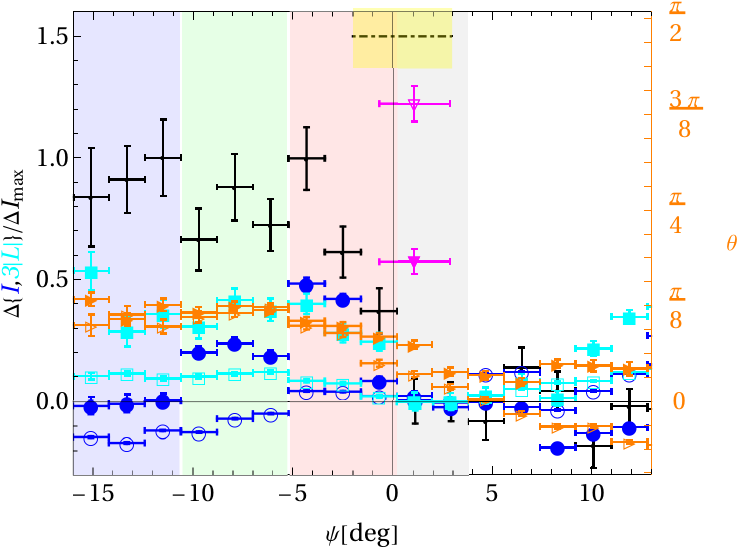}
            \put (37,75) {\small\textcolor{black}{Southwest sector}}
        \end{overpic}
    \end{center}
    \vspace{-0.5cm}
    \caption{\label{fig:EdgePsi}
   Radial profiles ($1\sigma$ error bars) of \gama-ray (small black disks), $30\GHz$ synchrotron (filled symbols), and $353\GHz$ dust (empty symbols) emission, as a function of oriented angular distance $\psi$ outside the FB edge, in the four (labeled) sectors.
   The total ($I$; circles) and linearly polarized ($|L|$; cyan squares, multiplied by $3$ for visibility) brightness excess above the foreground (based on the upstream, gray-shaded region), shown normalized to peak total brightness, rise abruptly downstream ($\psi\lesssim 0$); color shades define near (red), mid (green), and far (blue) downstream regions.
   The polarization angle $\theta$ (orange right-triangles, in \emph{WMAP} conventions) trend changes near the $\psi\simeq 0$ edge; the angle of the $\delta L$ jump across the edge (magenta down-triangles) is closer to the shock-normal angle $\theta_n$ (dot-dashed black line with yellow-shaded sector dispersion), especially in the west.
   	}
\end{figure*}

\begin{figure*}[h!]
    \begin{center}
        \hspace{-0.3cm}
        \begin{overpic}[width=0.44\linewidth]{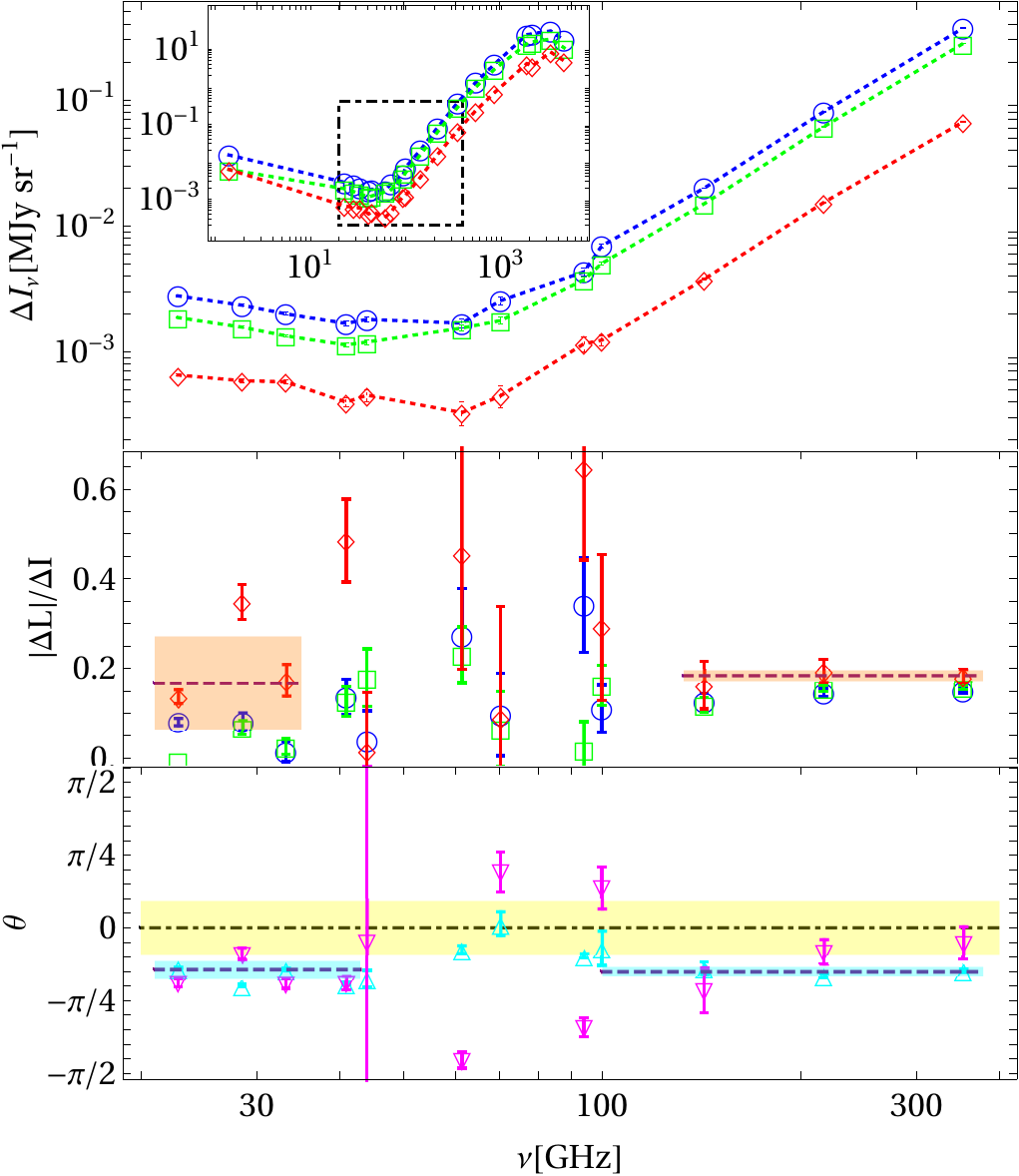}
            \put (40,102) {\small\textcolor{black}{North sector}}
        \end{overpic}
        \hspace{0.7cm}
        \begin{overpic}[width=0.44\linewidth]{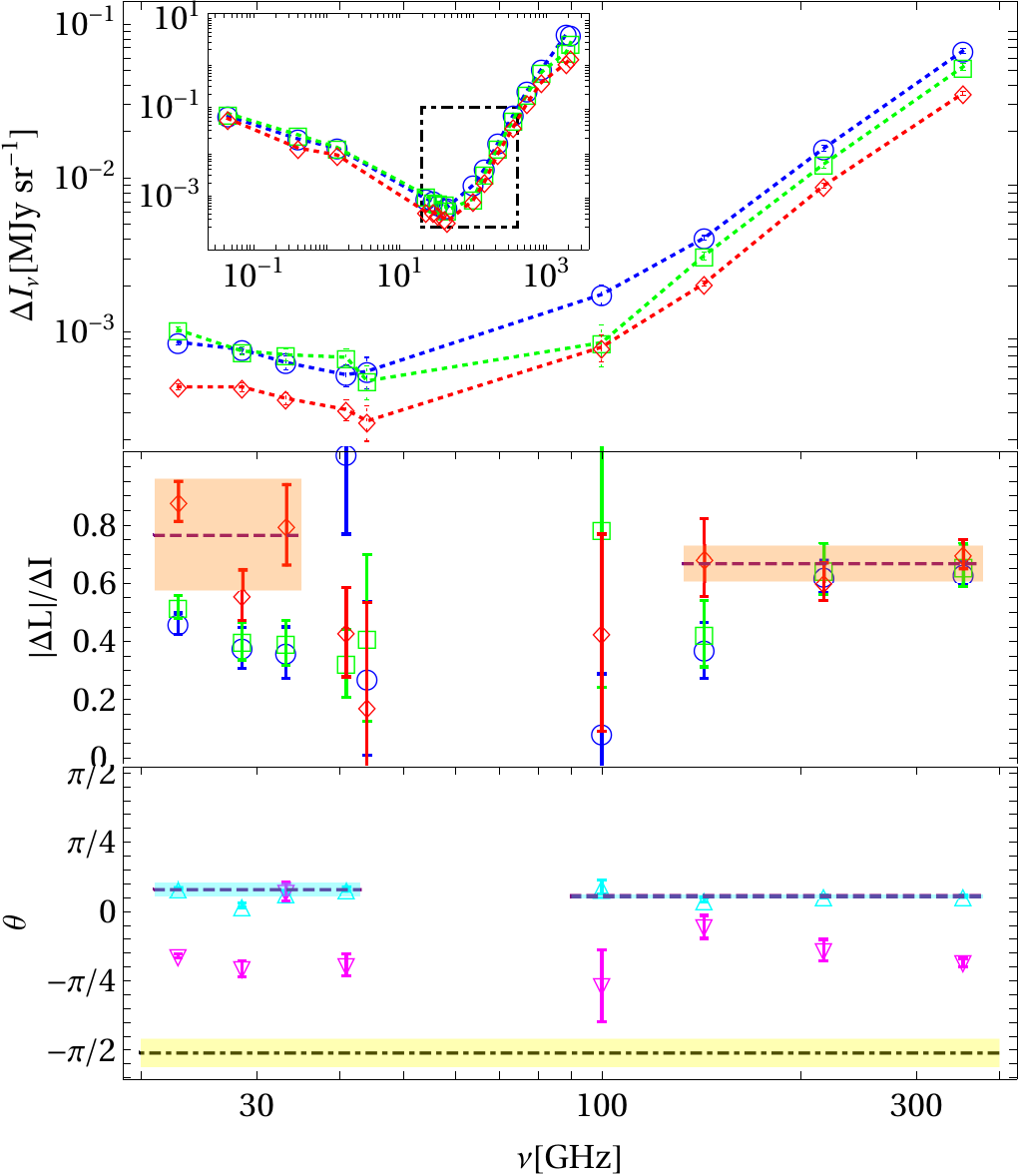}
            \put (37,102) {\small\textcolor{black}{Northwest sector}}
        \end{overpic}
    \end{center}
    \vspace{0.1cm}
    \begin{center}
        \hspace{-0.3cm}
        \begin{overpic}[width=0.44\linewidth]{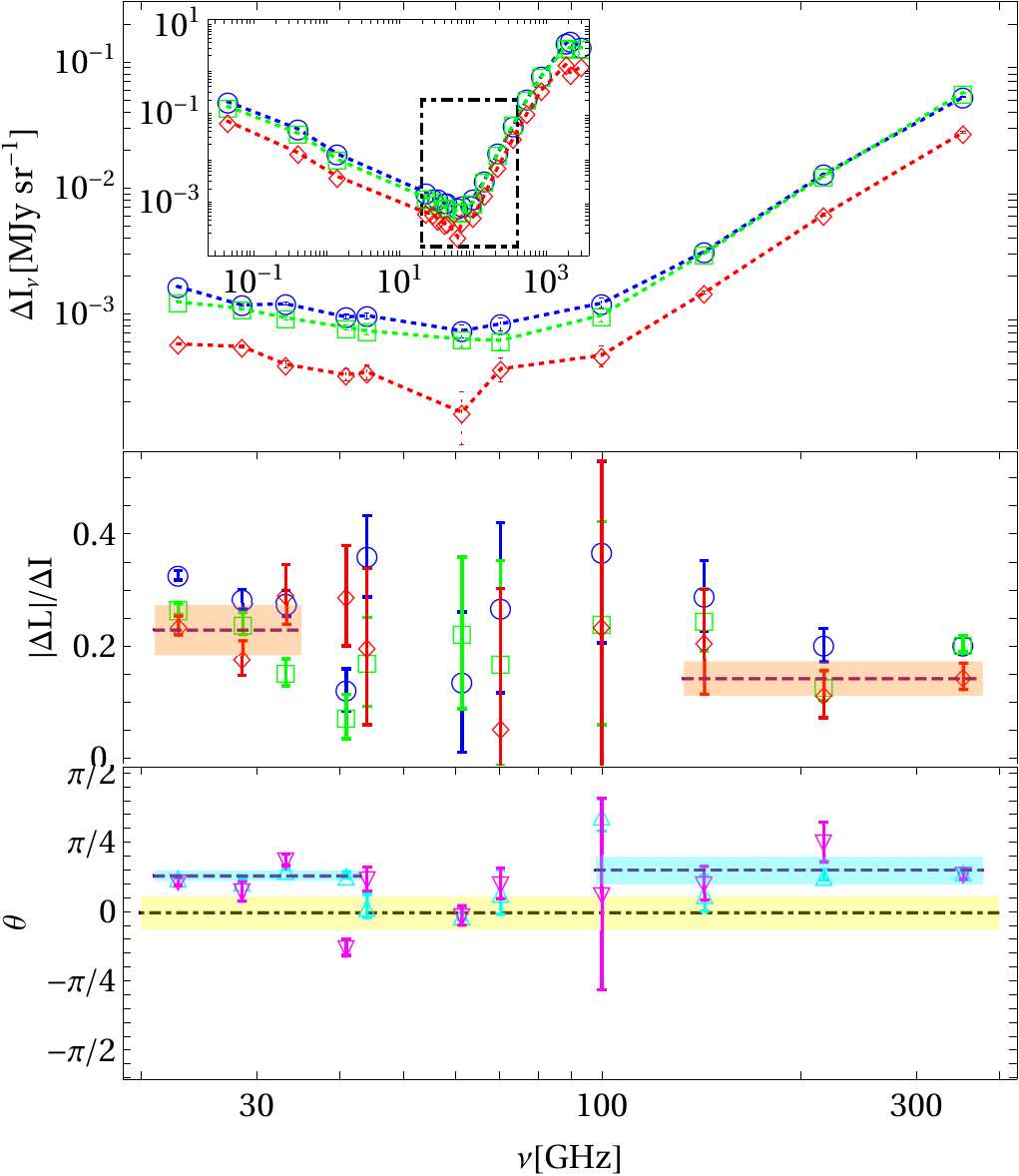}
            \put (40,102) {\small\textcolor{black}{South sector}}
        \end{overpic}
        \hspace{0.7cm}
        \begin{overpic}[width=0.44\linewidth]{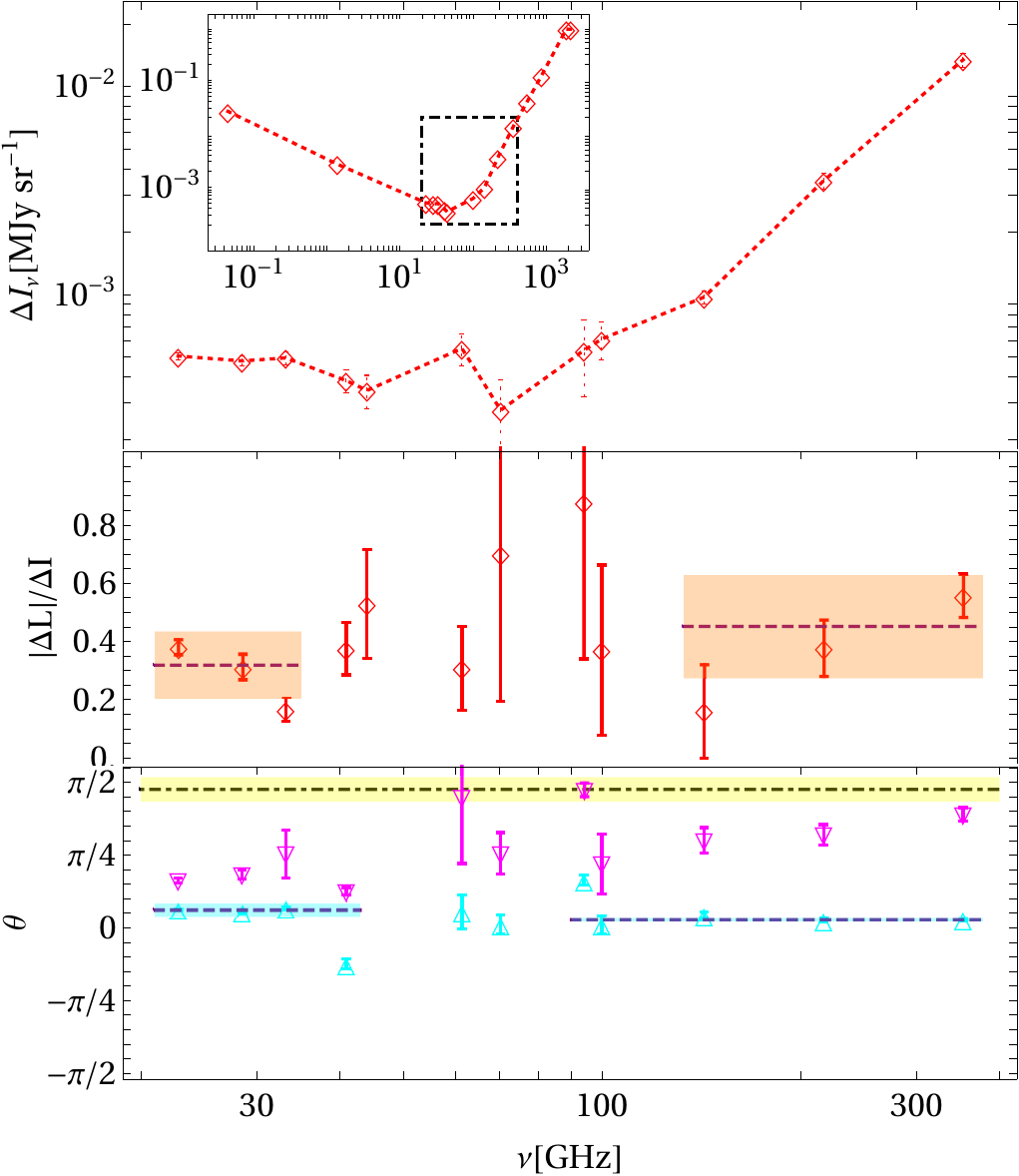}
            \put (37,102) {\small\textcolor{black}{Southwest sector}}
        \end{overpic}
    \end{center}
    \vspace{-0.7cm}
	\caption{\label{fig:Spect}
    Spectral analysis, in each sector (labeled panel composite), of the total brightness (top panels, with broader frequency-range insets as in {\UK}) and the linear polarization fraction (middle panels) of the excess in the near (red diamonds), mid (green squares) and far (blue circles) downstream regions, and of the polarization angles (bottom panels) $\theta_u$ upstream (cyan up triangles) and $\theta(\delta L)$ across the edge (magenta down triangles), compared to the edge-normal $\theta_n$ (dot-dashed line with yellow-shaded dispersion).
    The upstream and downstream regions are defined in Fig.~\ref{fig:EdgePsi}.
    The mean near-downstream $|\Delta L|/\Delta I$ and upstream $\theta_u$ are similar in the synchrotron and dust regimes, shown (dashed lines with shaded dispersions) separately.
    }
\end{figure*}

We search for the linearly polarized signal from the FBs at high, $|b|>30^\circ$ latitudes, where depolarization effects and confusion with Galactic structures diminish.
The FB outskirts, where polarization is expected, are examined in four different sectors, marked (by dashed white boundaries) in Fig.~\ref{fig:RGB}.
In each hemisphere, we choose one sector crossing the high-latitude tip of the bubble, and one sector on its west side; the eastern FB edges are avoided due to possible confusion with the putative lobe edges.
Although wide sectors would provide a better signal-to-noise ratio, we adopt fairly narrow sectors, along which the projected FB edge is approximately linear, in order to better pick up the polarized signal with a meaningful polarization angle.
Namely, each of the four chosen sections of the FB edges are nearly straight, with a $\sim$constant angle $\theta_e$ with respect to Galactic cardinal directions, so one can distinguish between a polarization angle $\theta$ parallel ($|\theta-\theta_e| \simeq \{0,\pi\}$) vs. perpendicular to the edge. Once an edge section is chosen, its associated sector is entirely dictated by the distance transform.

Figure \ref{fig:RGB} summarizes, for each sector (in its respective corner, using a Cartesian projection), the local FB edge morphology, its approximate orientation, and the inferred directions of the projected upstream (dashed cyan) and downstream excess (solid magenta) magnetic fields, based on both thermal dust (marked by circles) and synchrotron emission.
The FB polarization properties in each sector, including these inferred magnetic field orientations, are derived from the radial (Fig.~\ref{fig:EdgePsi}) and spectral (Fig.~\ref{fig:Spect}) profiles of the CMB-subtracted brightness $I$ in \emph{WMAP} and \emph{Planck} full-mission maps (\UK), along with their $Q$ and $U$ counterparts. 

Figure \ref{fig:EdgePsi} demonstrates the radial profiles of \gama-ray and microwave signals measured near the FB edge in each of the four sectors, stacked at different angular distances $\psi$ from the edge.
Here, $\psi$ increases outward, away from the GC, so the shock upstream (downstream) lies at $\psi\gtrsim0$ ($\psi\lesssim0$).
Foregrounds typically vary slowly with $\psi$ at the high latitudes and large angular scales of interest, and so can be approximated using their mean upstream value; the figure shows differences (denoted by $\Delta$; circles and squares) with respect to such fixed foreground estimates.
Using linearly-fitted foregrounds does not substantially change the results.

The brightness $I$ was previously found to abruptly rise as one crosses the FB edge downstream, in \gama-rays (\KG), X-rays \citep{Keshetgurwich18}, radio, microwaves, and infrared (\UK).
Figure \ref{fig:EdgePsi} shows a similar rise in the linearly-polarized microwave amplitude $|L|$ (squares), in all four sectors, where
\begin{equation}
L\equiv Q+iU \fin
\end{equation}
The downstream rise in $|L|$ is seen in both synchrotron and dust emission, represented respectively by the lowest ($30\GHz$; filled symbols) and highest ($353\GHz$; empty symbols) \emph{Planck} polarization channels; the $\sim60\GHz$ transition between synchrotron and dust regimes was established in {\UK} and is reproduced below.
The downstream polarized excess $|\Delta L|$ roughly follows the respective brightness excess $\Delta I$, and is found to be associated with the FBs at a TS-test confidence level exceeding $5\sigma$ in each sector and in each of these two channels, separately.
As the effect is well-localized and far from the radio-lobe edges, especially in the western sectors, we conclude that the localized $\Delta L$ excess is associated with the FBs and not with the lobes.

Figure \ref{fig:Spect} presents the FB microwave excess in all \emph{WMAP} and \emph{Planck} polarization channels, spanning the $23<\nu<353$ GHz frequency range.
For each sector, the $\Delta I$ (top panels) and $\Delta L$ (middle and bottom panels) properties are shown in the near (red diamonds), mid (green squares) and far (blue circles) downstream regions, which are defined (as shaded regions) in Fig.~\ref{fig:EdgePsi}.
The top panels and insets confirm that the $\nu\lesssim50\GHz$ and $\nu\gtrsim 100\GHz$ ranges are dominated respectively by $\Delta I\propto \nu^{\alpha}$ synchrotron and $\Delta I\propto \nu^{7/2}/(e^{h\nu/k_B T}-1)$ thermal dust emission.
Here, $\alpha\sim -0.8$ is the synchrotron spectral index, $T\sim30\mbox{ K}$ is the dust temperature, $h$ is Planck's constant, and $k_B$ is Boltzmann's constant; for a spectral analysis, see {\UK}.

In both synchrotron and dust regimes, we find (middle panels) a near-downstream linear polarization fraction
\begin{equation}\label{eq:AmplitudeL}
  |\Delta L|/\Delta I\simeq 20\% \coma
\end{equation}
in both north and south sectors, in which the foreground is fairly constant and robustly determined.
As the figure shows, the statistical errors are small in the synchrotron and dust regimes, but substantial in the intermediate, $\sim 60$--$90\GHz$ frequency range.
Averaging the non-intermediate channels over both sectors gives a polarization fraction $(19\pm5)\%$.
In the western sectors, the foreground and in particular its $|L(\psi)|$ component vary more rapidly with $\psi >0$, so our constant foreground approximation is less robust; this systematic error is especially large in the SW sector, where even $I$ can be estimated only in the near-downstream region.
Consequently, averaging the non-intermediate channels over all four sectors yields a more noisy, $(23\pm14)\%$ estimate of the fractional polarization. The dispersion among channels and sectors reflects systematic uncertainties of order $\sim2$ in polarization fraction, governed by the imprecise edge localization and foreground determination. Moreover,
the estimated values of $|\Delta L|/\Delta I$ are all projected, implying a higher intrinsic polarization fraction.

Throughout the analyzed region, the polarization angle
\begin{equation}\label{eq:theta}
  \theta(L) \equiv \arg(L)/2
\end{equation}
(right-triangles in Fig.~\ref{fig:EdgePsi}) lies in the $|\theta|\lesssim30^\circ$ range, with implied magnetic fields roughly following the morphology of the extended lobes, which dominate the polarized microwaves.
However, small changes in the $\theta(\psi)$ behavior are identified at the FB edge in all four sectors: small but significant $\theta$ jumps at the two western sectors, and $\theta(\psi)$ trend reversals in the other two sectors.
While the upstream $\theta_u$ (up-triangles in Fig.~\ref{fig:Spect}) can be determined robustly, it is difficult to measure the downstream angle $\theta_d$ of FB-associated emission, due to the strong, $\theta_f\simeq \theta_u$ foreground and off-edge projection effects.
Indeed, either overestimating or underestimating the foreground would offset the measured $\theta_d$ towards $\theta_f$, and projection effects dilute the anticipated, edge-perpendicular polarization.

In $Q$--$U$ space, the $L(\psi)$ trajectory shows a small $\delta L$ jump at $\psi\simeq0$ in all four sectors, thus providing a local, albeit noisy, estimate of the FB polarization angle with minimal projection effects.
The corresponding $\theta(\delta L)$ angle, presented (in magenta) in Figs.~\ref{fig:RGB} (solid lines show the inferred magnetic field orientation), \ref{fig:EdgePsi} and \ref{fig:Spect} (down-triangles), lies approximately half-way between the upstream $\theta_u$ and the shock normal, $\theta_n$ (black dot-dashed lines in Figs.~\ref{fig:EdgePsi}--\ref{fig:Spect}).
This effect, more noticeable in the western sectors where $\theta_u$ and $\theta_n$ are farther apart, indicates that the near-downstream magnetic field is indeed aligned preferentially parallel to the shock front.

\section{Discussion}

We examine the high-latitude tip section and a western intermediate-latitude section of the pre-determined edge of each of the FBs, chosen where the edge is approximately linear ($\theta_e\simeq\mbox{const.}$), and stack broadband data in bins parallel to these sections.
A polarized component of both synchrotron and dust emission is detected and associated ($>5\sigma$ in each section and each channel) with the FBs, thus separating their emission from the stronger polarized foreground arising from the more extended radio lobes.
A fractional $\sim20\%$ polarization is inferred for the projected FB signal near the tip of the bubbles, where the edge is robustly localized and foreground is better behaved, but a systematic uncertainty of order $\sim2$ should be taken into account, as well as projection effects that dilute the intrinsic polarization.
The signal is robustly associated with a change in polarization angle, indicating a preferential magnetic field alignment parallel to the edge, although the precise polarization angle of the FB emission cannot be uniquely separated from the foreground.

Thanks to their large extent on the sky, the FBs provide unique constraints on diverse physical processes, ranging from cosmic-ray diffusion ({\eg} {\KG}) to dust-grain alignment.
The polarization angles of synchrotron and dust emission are similar to each other throughout the analyzed sectors, as often found in the magnetized interstellar medium; some differences are however seen, especially upstream of the FBs.
Dust emission is thought to be preferentially polarized along the long grain axis, radiatively torqued perpendicular to the ambient magnetic field \citep{DolginovMitrofanov76, DraineWeingartner97, HoangLazarian08, Lazarian07}. 
Our results constrain the alignment timescale as $<1\Myr$ (for $|\psi|=$ a few degrees downstream) under the extreme conditions at the FB outskirts: a highly ionized, hot ($\sim0.5\keV$ electron temperature), and dilute ($10^{-3}\cm^{-3}$ density) gas, exposed to anisotropic, strong starlight (\UK).

There are interesting similarities between the extended, polarized lobes and the more compact FBs, invoking previous claims that the two are complementary signatures of the same phenomenon. However, the above arguments, in particular the farther westward extent of the lobes reaching beyond the FB edges, indicate that the lobes are not associated with the supersonic FBs or the outflows that produced them. The polarized FB signature found in this study is clearly distinguished from the lobes, affirming this conclusion.


The $\gtrsim 40\%$ larger projected extent of the lobes, their large, $d\gtrsim 5.5\kpc$ distance from the Sun inferred from their $|b|\lesssim10^\circ$ depolarization \citep{CarrettiEtAl13}, the similar curvature of their ridges and the GC spur \citep{JonesEtAl12}, and the absence of localized lobe--FB collision signatures, indicate that they are large-scale structures emanating from the GC vicinity, and not objects lying in the FB background or foreground.
Our results thus imply that the lobes are older (and so, more curved, by an initial offset, Galactic winds, or upstream inhomogeneity) than the FBs, encompassing the latter and being in part processed by the FB forward shocks.

The large lobes were previously interpreted as arising from a slow, $\gtrsim100\Myr$ star-formation-driven GC outflow, possibly linked to the $\sim 3\times 10^7M_\odot$ star-forming molecular gas ring \citep{CarrettiEtAl13}.
However, their similarities to the FBs in terms of morphology, synchrotron signature, and $\gtrsim 10^{55}\erg$ inferred energy, indicate that they are more likely to be the remnants of a collimated outburst from the SMBH, resembling but predating the FB outburst.
If so, the lobes may well be the haze-like signature of the \emph{ROSAT}/\emph{eROSITA} \citep{Sofue00, PredehlEtAl20} bubbles: the larger, $b\gtrsim 80^\circ$ counterparts of the FBs.

\section*{Acknowledgements}
I am grateful to Ilya Gurwich for helpful discussions.
This research was supported by The Israel Science Foundation (Grant No. 2126/22).

\bibliographystyle{elsarticle-harv}
\bibliography{FermiBubbles}

\end{document}